\documentclass{article}
\usepackage{spconf,amsmath,graphicx,amsthm,amssymb}
\usepackage{cite} 
\usepackage{multirow}
\usepackage{algorithm}
\usepackage{algpseudocode}
\usepackage{colortbl}  
\usepackage{tabularx}
\usepackage[table,xcdraw]{xcolor}
\usepackage{booktabs}
\usepackage{subcaption}
\usepackage[normalem]{ulem}

\title{Metric-oriented Speech Enhancement using Diffusion \\Probabilistic Model}
\name{Chen Chen, Yuchen Hu, Weiwei Weng, Eng Siong Chng}
\address{School of Computer Science and Engineering, Nanyang Technological University, Singapore \\ CHEN1436@e.ntu.edu.sg}
\begin{document}
\maketitle
\begin{abstract}
Deep neural network based speech enhancement technique focuses on learning a noisy-to-clean transformation supervised by paired training data. However, the task-specific evaluation metric (e.g., PESQ) is usually non-differentiable and can not be directly constructed in the training criteria. This mismatch between the training objective and evaluation metric likely results in sub-optimal performance. To alleviate it, we propose a metric-oriented speech enhancement method (MOSE), which leverages the recent advances in the diffusion probabilistic model and integrates a metric-oriented training strategy into its reverse process. Specifically, we design an actor-critic based framework that considers the evaluation metric as a posterior reward, thus guiding the reverse process to the metric-increasing direction. The experimental results demonstrate that MOSE obviously benefits from metric-oriented training and surpasses the generative baselines in terms of all evaluation metrics.
\end{abstract}
\begin{keywords}
Diffusion probabilistic model, speech enhancement, reinforcement learning 
\end{keywords}
\section{Introduction}
\label{sec:intro}

Recent advances in deep learning has brought remarkable success to the speech enhancement technique, where a noisy-to-clean transformation is learned to remove additive noises by a supervised learning manner~\cite{wang2018supervised,xu2014regression,koizumi2020speech,chen2021time}. However, this paradigm suffers from a mismatch between training and evaluation: the training criterion (e.g., Mean Square Error) must be differentiable for gradient calculation~\cite{chen2015speech}, while the evaluation metric (e.g. PESQ) are usually non-differentiable, thus can not be directly modeled in loss function as minimized objective. Consequently, the optimized model after training can not achieve best performance in terms of evaluation metric. \par
This mismatch is also reported in other supervised learning tasks, such as machine translation~\cite{bahdanau2016actor,wu2018study} and automatic speech recognition~\cite{prabhavalkar2018minimum,chen2022self,chen2022leveraging}. Prior works have utilized reinforcement learning (RL) based algorithms to harmonize the mismatch using metric-based training approach~\cite{rennie2017self}, as these tasks contain a sequential decoding process that can be naturally viewed as Markov Decision Process (MDP)~\cite{tjandra2018sequence}. Nevertheless, as a regression task, mainstream SE approaches train a one-shot discriminative model without the time-step concept for MDP, which is infeasible for RL-based optimization. \par        
Diffusion probabilistic model~\cite{ho2020denoising}, showing outstanding results in generative tasks~\cite{nichol2021improved, kong2020diffwave}, brings possibility for metric-based optimization of SE task, as it inherently consists of MDP-based diffusion and reverse processes~\cite{luo2021diffusion}. More specifically, an isotropic Gaussian distribution is added to the clean speech during step-by-step diffusion process, and in the reverse process, gradually estimates and subtracts additive noise to restore the clean input~\cite{lu2021study}. \par
In this work, we present a metric-oriented speech enhancement method called MOSE, which effectively constructs the non-differentiable metric into the training objective. Inspired by actor-critic based algorithm~\cite{chen2019towards}, we design a value-based neural network that is updated by Bellman Error~\cite{grondman2012survey} to evaluate current policy in terms of metric-related reward function, then it guides the prediction of subtracted noise in a reverse process by the differentiable manner. In this way, the original policy is optimized to the metric-increasing direction, while the value-based network is trained to provide reasonable feedback. Experimental results demonstrate that MOSE obviously benefit from metric-oriented training and beat other generative methods in terms of all metrics. Furthermore, it shows better generalization in face of unseen noises with large domain mismatch.

\begin{figure*}[t]
\centering
\includegraphics[width=0.92\textwidth]{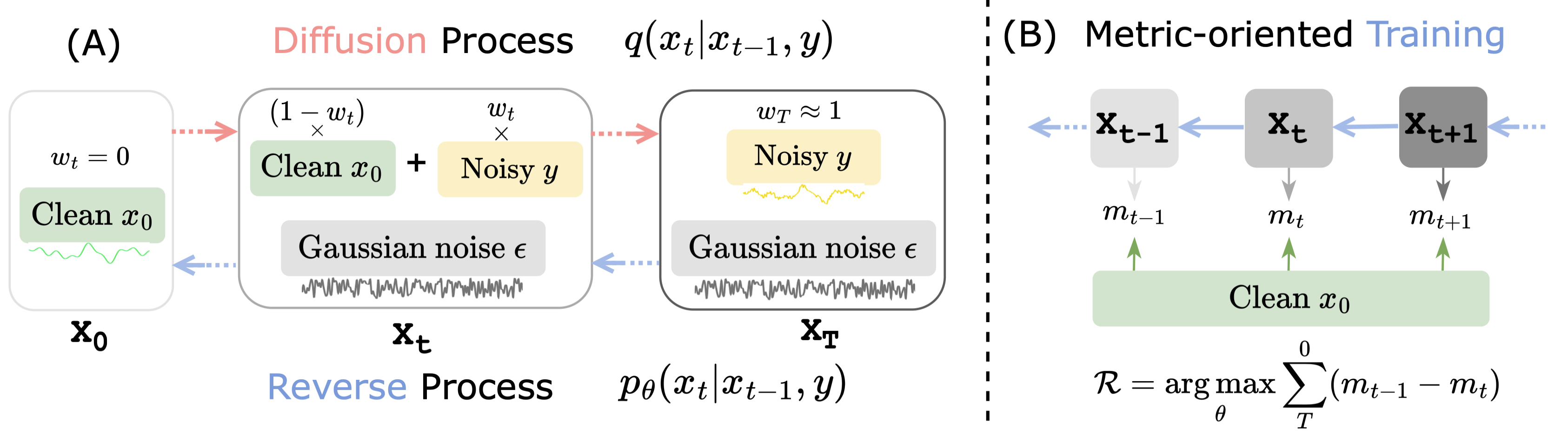}
\vspace{-0.2cm}
\caption{The conditional diffusion probabilistic model (A) and metric-oriented training (B). The red and blue arrows respectively denotes the diffusion and reverse process. $w_t$ is the weight of linear interpolation, and $m_t$ is the task-specific metric.}
\vspace{-0.4cm}
\label{f1}
\end{figure*}

\section{Preliminaries}
We first define noisy speech as $y$ and define its corresponding ground-truth clean speech as $x_0$. The speech enhancement task aims to learn a transformation $f$ that converts the noisy input to clean signal: $x_0 = f(y),\ x_0, y \in \mathbb{R}^L$. 
\subsection{Diffusion Probabilistic Model}\label{2.1}
In this part, we briefly introduce the diffusion process and the reverse process of the typical diffusion probabilistic model. \par
\noindent\textbf{Diffusion process} is formulated as a $T$-step Markov chain that gradually adds Gaussian noise to the clean signal $x_0$ in each step $t$. The Gaussian model is denoted as $q(x_t|x_{t-1}) = \mathcal{N}(x_t;$ $\sqrt{1-\beta_{t}}x_{t-1},\beta_{t}I)$, where $\beta_{t}$ is a small positive constant that serve as a pre-defined schedule. With enough diffusion step $T$, the latent variable $x_T$ can be finally converted to an isotropic Gaussian distribution $p_{latent}(x_T)=\mathcal{N}(0,I)$. Therefore, based on $x_0$, the sampling distribution of each step in the Markov chain can be derived as the following:
\begin{equation}
  q(x_t|x_0) = \mathcal{N}(x_t;\sqrt{\bar{\alpha}_t}x_0,(1-\bar{\alpha}_t)I),
  \label{eq1}
\end{equation}
where $\alpha_t = 1-\beta_t$ and $\bar{\alpha}_t =  \prod_{s=1}^{t} \alpha_s$. \par
\noindent\textbf{Reverse process} aims to restore the $x_0$ from the latent variable $x_T$ along another Markov chain, which is denoted as $p_{\theta}(x_{t-1} | x_t)$, where $\theta$ is learnable parameters. As marginal likelihood  $p_{\theta}(x_0) = \int p_\theta(x_0, \cdots, x_{T-1}|x_T)\cdot p_{\text{latent}}(x_T) dx_{1:T}$ is intractable for calculation, the ELBO~\cite{ho2020denoising} is utilized to approximate a learning objective for neural model training. Therefore, the equation of the reverse process can be denoted as:
\vspace{-0.2cm}
\begin{equation}
\begin{split}
\begin{aligned}
 p_{\theta}(x_{t-1}|x_t) &= \mathcal{N}(x_{t-1};\mu_\theta(x_t,t),\tilde{\beta}_t I), \\
 \text{where} \quad \mu_\theta(x_t,t) &= \frac{1}{\sqrt{\alpha_t}}(x_t-\frac{\beta_t}{\sqrt{1-\bar{\alpha}_t}}\epsilon_\theta(x_t,t))
 \label{eq2}
\end{aligned}
\end{split}
\end{equation} 
Here $\mu_{\theta}(x_t,t)$ denotes the mean of $x_{t-1}$, which is obtained by subtracting the estimated Gaussian noise $\epsilon_{\theta}(x_t,t)$ in the $x_t$. Furthermore, the variance is derived to a constant $\tilde{\beta}_t = \frac{1-\bar{\alpha}_{t-1}}{1-\bar{\alpha}_t}\beta_t$.

\subsection{Reinforcement Learning}\label{2.2}
Reinforcement learning (RL) is typically formulated as a Markov Decision Process (MDP) that includes a tuple of trajectories $\left \langle \mathcal{S},\mathcal{A},\mathcal{R}, \mathcal{T} \right \rangle$. For each time step $t$, the agent considers state $s_t \in S$ to generate an action $a_t \in \mathcal{A}$ which interacts with environment. The transition dynamics $\mathcal{T}(s_{t+1}|s_t,a_t)$ is defined as transition probability from current state $s_t$ to next state $s_{t+1}$, and gain an instant reward $r_t(s_t,a_t)$. The objective of RL is to learn optimal policy to maximize the cumulative reward $\mathcal{R}$ along all time steps.\par
Since the diffusion probabilistic model formulates speech enhancement task as MDP in section~\ref{2.1}, the RL algorithm can be integrated in the reverse process to explore optimal policy. More specifically, given the current state $x_t$, the policy network is supposed to predict a Gaussian noise $\epsilon_t$ as the current action. After subtracting the $\epsilon_t$ in $x_t$, the $x_{t-1}$ is obtained as next state, as step number $t$ is decreasing during reverse process. Furthermore, the instant reward $r_t$ is calculated by comparison of $x_t$ and $x_{t-1}$, which guides the update of parameters $\theta$ during model training.

\section{Methodology}
In this section, we introduce our proposed MOSE, which integrates the metric-oriented training into the reverse process of a conditional diffusion probabilistic model. The overview of MOSE is shown in Fig.~\ref{f1}.

\subsection{Conditional Diffusion Probabilistic Model}
As real-world noises usually does not obey the Gaussian distribution, we incorporate noisy speech $y$ into the procedures as a conditioner in this part. Specifically, a dynamic weight $w_t\in[0,1]$ is employed for linear interpolation from $x_0$ to $x_T$. Therefore, as shown in Fig.~\ref{f1}, each latent variable $x_t$ consists of three parts: clean component $(1-w_t)\times x_0$, noisy component $w_t\times y$, and Gaussian Noise $\epsilon$. Furthermore, the diffusion process in Eq.~(\ref{eq1}) can be rewritten as:   
\begin{align}
  & q(x_t|x_0,y) = \mathcal{N}(x_t;(1-w_t)\sqrt{\bar{\alpha}_t}x_0 + w_t\sqrt{\bar{\alpha}_t}y ,\delta_tI), \\
  &\text{where} \quad \ \delta_t = (1-\bar{\alpha}_t)-w_t^2\bar{\alpha}_t
\label{eq4}
\end{align}\par
The conditional reverse process starts from $x_T$ with $w_T=1$, which is denoted as $\mathcal{N}(x_T,\sqrt{\bar{\alpha}_T}y,\delta_TI)$. Referring to Eq.~(\ref{eq2}), we denoted the conditional reverse process as:
\begin{equation}
  p(x_{t-1}|x_t,y) = \mathcal{N}(x_{t-1};\mu_\theta(x_t,y,t),\tilde{\delta}_tI),
  \label{eq5}
\end{equation}
where $\mu_{\theta}(x_t,y,t)$ is the predicted mean of variance $x_{t-1}$. It means that the neural model $\theta$ considers both variance $x_t$ and noisy conditioner $y$ during its prediction. Therefore, similar to Eq.~(\ref{eq2}), we define the mean of $\mu_{\theta}$ as a linear combination of $x_t$, $y$, and $\epsilon_{\theta}$:
\vspace{-0.2cm}
\begin{equation}
  \mu_\theta(x_t,y,t) = c_{xt}x_{t} + c_{yt} y - c_{\epsilon t} \epsilon_\theta(x_t,y,t),
  \label{eq6}
\end{equation}
where the coefficients $ c_{xt}, c_{yt}, $ and $ c_{\epsilon t} $ can be derived from the ELBO optimization criterion in~\cite{lu2022conditional}. Finally, we combine Gaussian noise $\epsilon$ and non-Gaussian noise $y-x_0$ as ground-truth $C_t^{noise}$:
\vspace{-0.2cm}
\begin{align}
&C_t^{noise}(x_0,y,\epsilon) = \frac{m_t\sqrt{\bar{\alpha}_t}}{\sqrt{1-\bar{\alpha}_t}}{(y-x_0)} + \frac{\sqrt{\delta_t}}{\sqrt{1-\bar{\alpha}_t}}\epsilon \label{eq10}\\ 
&\nabla_\theta\mathcal{L}_1 = \parallel C_t^{noise}(x_0,y,\epsilon) - \nabla_\theta \epsilon_\theta(x_t,y,t) \parallel_1 
\label{eq11}
\end{align}
where $C_t^{noise}$ provides supervision information, and $\mathcal{L}_1$ is calculated for back propagation of neural network.

\begin{figure}[t]
\centering
\includegraphics[width=0.43\textwidth]{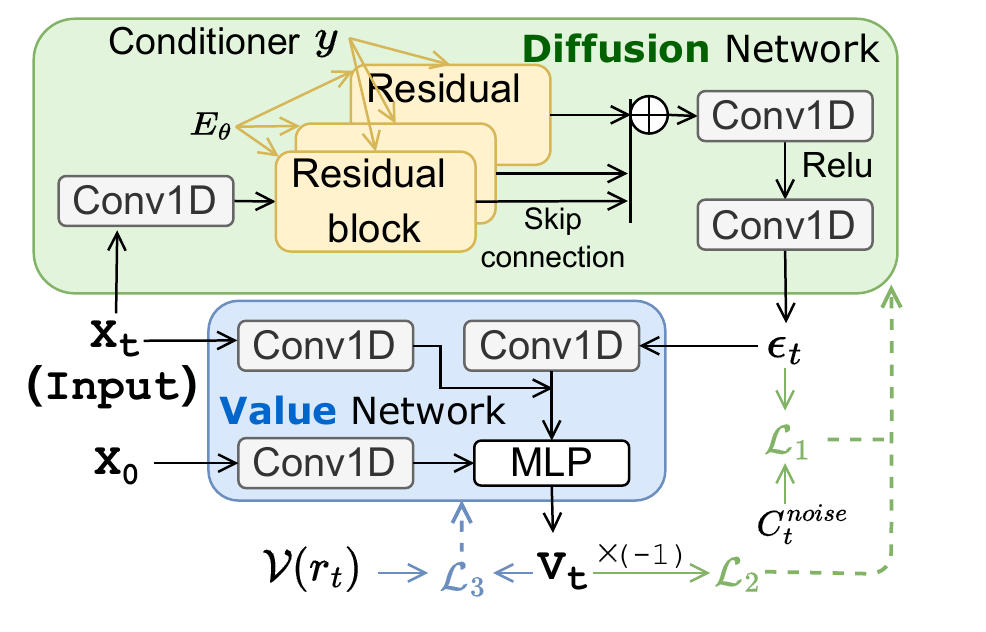}
\vspace{-0.3cm}
\caption{The main structure of MOSE. Dashed line stands for back propagation for neural network.}
\vspace{-0.5cm}
\label{f2}
\end{figure}

\subsection{Metric-oriented Training}
Given the task-specific evaluation metric $m$, each $t$-step variable can calculate the $m_t$ by $x_t$ and $x_0$ as they are in same shape. In order to directly optimize $m_t$, an actor-critic RL algorithm is integrated into conditional reverse process, as shown in the Fig.~\ref{f1} (B). \par
Since we hope that the latent variable is iterated toward the metric-increasing direction in the reverse process, the reward function is customized as: $r_t = m_{t-1}-m_{t}$, where $t$ starts from $T$ to 0. However, posterior $r_t$ is obviously non-differentiable for $\theta$, thus failing to propagate gradient. To this end, we further employ a \textbf{Value network $V$} with parameter $\theta_v$ as the blue box in Fig.~\ref{f2}, and the original network is denoted as \textbf{Diffusion network $D$} with parameter $\theta_d$ for distinction. In general, The Diffusion network consumes $x_t$ to predict the subtracted noise $\epsilon_t$ as action, while the Value network generates an score $v_t$ to evaluate this $\epsilon_t$ based on $x_t$. The training strategy of MOSE is explained in Algorithm~\ref{algo1}.\par
\begin{algorithm}[t]
  \caption{MOSE Training}\label{algo1}
  \begin{algorithmic}[1]
    \State Randomly initialize the Diffusion network $D(x |\theta_d)$ and Value network $V(x,\epsilon |\theta_v)$.
    \State Initialize $N_{total}$, $N_{th}$, $\gamma$, and $\alpha$
    \For{$i=1,2,\cdots,N_{total}$}
    \State {Sample $(x_0,y)$ from Dataset} 
    \State {Sample $\epsilon{\sim}\mathcal{N}(0,I)$ and $t{\sim}\text{Uniform}(\{1,\cdots,T\})$}
    \State {Set $x_t = ((1-m_t)\sqrt{\bar{\alpha}_t}x_0 + m_t\sqrt{\bar{\alpha}_t}y) + \sqrt{\delta_t}\epsilon$ } 
    \State {Calculate $C_t^{noise}$ according to Eq.~(\ref{eq10})} 
    \If{$i<N_{th}$}
    \State {Update network $D$ by minimizing $\mathcal{L}_1$ in Eq.~(\ref{eq11})}
    \Else
    \State {Calculate $\epsilon_t = D (x_t,y,t|\theta_d)$ as action}
    \State {Calculate $\nabla_{\theta_d} \mathcal{L}_2 = - V (x_t,\nabla_{\theta_d}\epsilon_t, x_0|\theta_v)$}
    \State {Update $D$ by minimizing $\mathcal{L} = \mathcal{L}_1 + \alpha \cdot \mathcal{L}_2$}
    \State {Calculate $x_{t-1}$ according to Eq.~\ref{eq6} as next state}
    \State {Calculate $r_t = m_{t-1}(x_{t-1},x_0) - m_t(x_{t-1},x_0) $}
    \State {Set $\mathcal{V}_{t}= r_t + \gamma V(x_{t-1}, D(x_{t-1},y,t-1),  x_0|\theta_v)$ }
    \State {Calculate $\nabla_{\theta_v}\mathcal{L}_3= (\mathcal{V}_{t}-\nabla_{\theta_v}V(x_t,\epsilon_t, x_0|\theta_v))^2$}
    \State {Update network $V$ by minimizing $\mathcal{L}_3$}
    \EndIf
    \EndFor
  \end{algorithmic}
\label{algo1}
\end{algorithm}
MOSE starts training with conventional ELBO optimization, as explained from line 3$\sim$9 in Algorithm~\ref{algo1}, only Diffusion network $D$ is trained for $N_{th}$ iterations. Then we present joint training of Diffusion network $D$ and Value network $V$ from lines 10$\sim$18. Minimizing $\mathcal{L}_2=- V(x_t, \epsilon_t,x_0|\theta_v)$ indicates that $D$ tents to gain higher score from $V$, and $\mathcal{L}_2$ is simultaneously incorporated with a weight $\alpha$ to stabilize training. In order to encourage Value network $V$ to provide reasonable evaluation, we employ widely used Bellman Error~\cite{grondman2012survey} (line 17) to update $V$, where $\gamma$ is a decay factor for future reward. Consequently, the output score $v_t$ both considers current and future rewards based on the task-specific metric. For inference, we adopt a fast sampling scheme as same as in~\cite{kong2020diffwave}.
\section{Experiment}
\subsection{Experimental Setup}
\noindent\textbf{Database}. We choose the publicly available VoiceBank DEMAND dataset~\cite{valentini2016investigating} for SE training and evaluation. Specifically, the training set contains 11,572 noisy utterances from 28 speakers and is mixed by 10 different types with four SNR levels (0, 5, 10, and 15 dB) at a sampling rate of 16 kHz, as well as their corresponding clean utterances. The test set contains 5 types of unseen noise in SNR levels (2.5, 7.5, 12.5, and 17.5 dB). To evaluate the performance of a model in unseen noises, we further mix the test set of TIMIT~\cite{garofolo1988getting} and ``helicopter" and ``babycry" noises with different SNR levels (-6, -3, 0, 3, 6 dB), where a large domain mismatch exists between training and testing. \par
\noindent\textbf{Configuration}. The internal structure of MOSE is shown in Fig.~\ref{f2}. We employ 30 residual blocks with 64 channels in Diffusion Net. MLP block contains 4 linear layers with ReLU activation function. For training, MOSE takes 50 diffusion steps with training noise schedule $\beta_t \in [1\times10^{-4},0.035]$, and the interpolation weight $m_t = \sqrt{(1- \bar{\alpha}_t)/\sqrt{\bar{\alpha}_{t}}}$. The $N_{total}$, $N_{th}$, and $\gamma$ in Algorithm~\ref{algo1} are respectively set as 40k, 30k, and 0.95. The initial learning rate of Diffusion network is set as $2\times10^{-4}$ for first $N_{th}$ iterations, and decrease to $1\times10^{-4}$ for $N_{th}\sim N_{total}$ iterations. The learning rate of the Value network is set as $1\times10^{-5}$. Both networks are optimized by Adam with a batch size of 32. The fast sampling method keeps the same schedule with~\cite{lu2022conditional}.  \par
\noindent\textbf{Metric}. We select the perceptual evaluation of speech quality (PESQ) as the task-specific metric of optimization objective due to its universality. Furthermore, prediction of the signal distortion (CSIG), prediction of the background intrusiveness (CBAK), and prediction of the overall speech quality (COVL) are also reported as references.
\begin{figure}[t]
\centering
\includegraphics[width=0.48\textwidth]{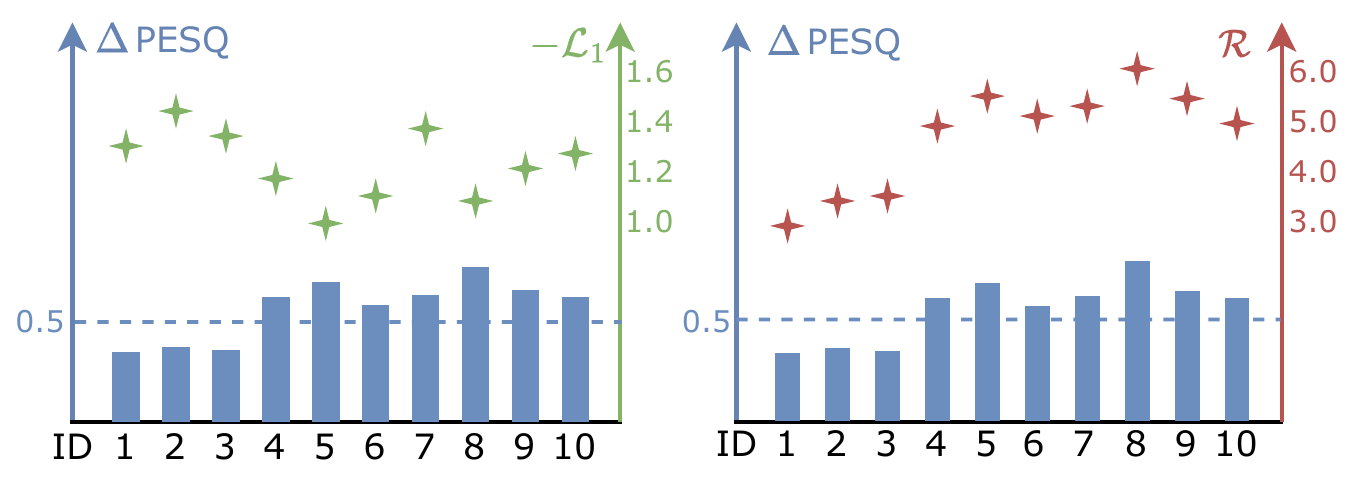}
\vspace{-0.8cm}
\caption{The relationship between $\Delta$PESQ and training loss $-\mathcal{L}_1$, as well as gained reward $\mathcal{R}$.}
\vspace{-0.2cm}
\label{f2}
\end{figure}
\vspace{-0.2cm}
\subsection{Result and Analysis}
\begin{table}[t]
\caption{Result of metric-oriented training.}
\vspace{-0.2cm}
\centering
\resizebox{0.45\textwidth}{!}{%
\begin{tabular}{c | c | c| c c c c c}
\toprule[1.2pt]
 ID &System                & $\alpha$ & PESQ & CSIG & CBAK & COVL &    \\ \midrule[1.2pt]
1 &  Unprocessed            &     -    & 1.97 & 3.35 & 2.44 & 2.63 &   \\ \midrule
2 &MOSE              &     0    & 2.44 &  3.65  &  2.87 & 3.01     \\ \midrule

3& \multirow{3}{*}{MOSE}   & 0.1    &  2.48  & 3.66 & 2.90 & 3.06 \\
4&                         & 1      &  \textbf{2.54}  & \textbf{3.73} & \textbf{2.93} & \textbf{3.12}      \\
5&                         & 5      &  2.51  & 3.69 & 2.91 & 3.08    \\ \bottomrule[1.2pt]
\end{tabular}}
\vspace{-0.4cm}
\label{t1}
\end{table}

\begin{table}[t]
\caption{MOSE \emph{vs.} other methods. ``Gen." and ``Dis." respectively denote generative and discriminative models.}
\vspace{-0.2cm}
\centering
\resizebox{0.45\textwidth}{!}{%
\begin{tabular}{c | c| c c c c c}
\toprule[1.2pt]
System              & Type & PESQ & CSIG & CBAK & COVL &    \\ \midrule[1.2pt]
Unprocessed         & - & 1.97 & 3.35 & 2.44 & 2.63 &   \\ \midrule
DSEGAN~\cite{phan2020improving}      &  Gen. & 2.39 & 3.46 & 3.11 & 2.90  \\
SE-Flow~\cite{strauss2021flow}     &  Gen. & 2.28 & 3.70 & 3.03 & 2.97      \\ 
CDiffuSE~\cite{lu2022conditional}    &  Gen. & 2.52 & 3.72 & 2.91 & 3.10  \\ \midrule
WaveCRN~\cite{hsieh2020wavecrn}     &  Dis. & 2.64 & 3.94 & \textbf{3.37} & 3.29    \\ 
Conv-TasNet~\cite{luo2019conv} &  Dis. & \textbf{2.67} & \textbf{3.94} & 3.31 & \textbf{3.30}   \\\midrule
MOSE (ours)         &  Gen. & 2.54 & 3.72 & 2.93 & 3.06    \\
\bottomrule[1.2pt]
\end{tabular}}

\label{t3}
\end{table}

\begin{table}[t]
\vspace{-0.2cm}
\caption{PESQ results on TIMIT dataset with different SNRs. ``Avg" denotes the average of all SNR levels.}
\vspace{-0.1cm}
\resizebox{0.49\textwidth}{!}{\begin{tabular}{c|cccccl}
\toprule[1.2pt]
 \multirow{2}{*}{System}         &   \multicolumn{6}{c}{Noise level, SNR =}                  \\
  &  -6  &  -3         & 0         & 3    & 6  & Avg.        \\ \midrule[1.2pt] 
\multicolumn{7}{c}{\cellcolor[HTML]{E0E0E0}\emph{Noise type: Helicopter}} \\ 
Unprocessed  & 1.05  & 1.07 & 1.10 & 1.16 & 1.26 & 1.13 \ \textcolor{red}{+0\%} \\
Conv-TasNet~\cite{luo2019conv}   & 1.06  & 1.08 & 1.14 & 1.21 & \textbf{1.47} & 1.19 \ \textcolor{red}{+5.3\%}  \\
MOSE      & \textbf{1.08}  & \textbf{1.13} & \textbf{1.16} & \textbf{1.26} & 1.44 & \textbf{1.21} \ \textcolor{red}{+7.1\%} \\ \midrule
\multicolumn{7}{c}{\cellcolor[HTML]{E0E0E0}\emph{Noise type: Baby-cry}} \\ 
Unprocessed  & 1.06  & 1.09  & 1.13 & 1.18 & 1.27  & 1.15 \ \textcolor{red}{+0\%}  \\
Conv-TasNet~\cite{luo2019conv}   & 1.06  & 1.10  & 1.15 & 1.21 & 1.37  & 1.18 \ \textcolor{red}{+2.6\%}   \\
MOSE       & \textbf{1.08}  & \textbf{1.13}  & \textbf{1.16} & \textbf{1.24} & \textbf{1.45}  & \textbf{1.21} \ \textcolor{red}{+5.2\%} \\ \bottomrule[1.2pt]
\end{tabular}}
\vspace{-0.4cm}
\label{table3}
\end{table}

\subsubsection{Experimental validation of mismatch}
We first design an experiment to verify the mismatch problem between the training objective and evaluation metric, and illustrate how we mitigate it. To this end, we train a typical diffusion probabilistic model, where $\mathcal{L}_1$ in Eq~(\ref{eq11}) is set as the only training objective. Then we sample 10 utterances and add up their $\mathcal{L}_1$ (50 steps), as well as calculate the improvement of PESQ ($\Delta$PESQ). The comparison is visualized in the left part of Fig.~\ref{f2}, and we observe that there is no correlation between $\mathcal{L}_1$ and $\Delta$PESQ, which indicates that SE model trained only by $\mathcal{L}_1$ will lead to sub-optimal performance in terms of PESQ. Meanwhile, we calculate the cumulatively gained reward $\mathcal{R}$ of these utterances after metric-oriented training and visualize in the right of Fig.~\ref{f2}, where an obvious positive correlation can be observed between $\Delta$PESQ and $\mathcal{R}$.

\subsubsection{Effect of metric-oriented training}
We then examine the effect of proposed metric-oriented training, and the results are reported in Table~\ref{t1}. ``Unprocessed" denotes direct evaluation based on noisy data, and $\alpha$ is the weight of $\mathcal{L}_2$ in Algorithm~\ref{algo1}. When $\alpha=0$, SE model are only trained by $\mathcal{L}_1$ loss. We observe that system 3$\sim$5 all surpass system 2 with help of metric-oriented training. When $\alpha=1$, the SE model achieves the best performance. 


In addition, Table~\ref{t3} summarizes the comparison between MOSE and other competitive SE methods, which contains 3 generative models and 2 discriminative methods. We observe that MOSE surpasses generative baselines in terms of all metrics, however, the best performance is still achieved by discriminative method. \par
\vspace{-0.25cm}
\subsubsection{Generalization on unseen noise}
We evaluate our trained model in unseen noisy condition with a wide range of SNR levels, where Conv-TasNet method is reproduced for comparison. The PESQ results are shown in Table~\ref{table3}. Despite gaining outstanding performance on the matched test set, we observed that the PESQ of Conv-TasNet dramatically degrades due to noise domain mismatch. However, the MOSE performs better than Conv-TasNet in terms of PESQ, especially in low-SNR conditions.   

\section{Conclusion}
\vspace{-0.15cm}
In this paper, we propose a speech enhancement method, called MOSE, which addresses the mismatch problem between training objective and evaluation metric. The probabilistic diffusion model is leveraged as MDP based framework, where metric-oriented training is presented in the reverse process. The experimental results demonstrate that MOSE beats other generative baselines in terms of all metrics, and show better generalization on unseen noises.

\vfill\pagebreak

\newpage


\bibliographystyle{IEEEtran}
\bibliography{refs}

\begin{thebibliography}{10}
\providecommand{\url}[1]{#1}
\csname url@samestyle\endcsname
\providecommand{\newblock}{\relax}
\providecommand{\bibinfo}[2]{#2}
\providecommand{\BIBentrySTDinterwordspacing}{\spaceskip=0pt\relax}
\providecommand{\BIBentryALTinterwordstretchfactor}{4}
\providecommand{\BIBentryALTinterwordspacing}{\spaceskip=\fontdimen2\font plus
\BIBentryALTinterwordstretchfactor\fontdimen3\font minus
  \fontdimen4\font\relax}
\providecommand{\BIBforeignlanguage}[2]{{%
\expandafter\ifx\csname l@#1\endcsname\relax
\typeout{** WARNING: IEEEtran.bst: No hyphenation pattern has been}%
\typeout{** loaded for the language `#1'. Using the pattern for}%
\typeout{** the default language instead.}%
\else
\language=\csname l@#1\endcsname
\fi
#2}}
\providecommand{\BIBdecl}{\relax}
\BIBdecl

\bibitem{wang2018supervised}
D.~Wang and J.~Chen, ``Supervised speech separation based on deep learning: An
  overview,'' \emph{IEEE/ACM Transactions on Audio, Speech, and Language
  Processing}, vol.~26, no.~10, 2018.

\bibitem{xu2014regression}
Y.~Xu, J.~Du, L.-R. Dai, and C.-H. Lee, ``A regression approach to speech
  enhancement based on deep neural networks,'' \emph{IEEE/ACM Transactions on
  Audio, Speech, and Language Processing}, vol.~23, no.~1, 2014.

\bibitem{koizumi2020speech}
Y.~Koizumi, K.~Yatabe, M.~Delcroix, Y.~Masuyama, and D.~Takeuchi, ``Speech
  enhancement using self-adaptation and multi-head self-attention,'' in
  \emph{ICASSP 2020-2020}, 2020.

\bibitem{chen2021time}
C.~Chen, N.~Hou, D.~Ma, and E.~S. Chng, ``Time domain speech enhancement with
  attentive multi-scale approach,'' in \emph{2021 Asia-Pacific Signal and
  Information Processing Association Annual Summit and Conference (APSIPA
  ASC)}.\hskip 1em plus 0.5em minus 0.4em\relax IEEE, 2021, pp. 679--683.

\bibitem{chen2015speech}
Z.~Chen, S.~Watanabe, H.~Erdogan, and J.~R. Hershey, ``Speech enhancement and
  recognition using multi-task learning of long short-term memory recurrent
  neural networks,'' in \emph{Conference of the International Speech
  Communication Association}, 2015.

\bibitem{bahdanau2016actor}
D.~Bahdanau, P.~Brakel, K.~Xu, A.~Goyal, R.~Lowe, J.~Pineau, A.~Courville, and
  Y.~Bengio, ``An actor-critic algorithm for squence prediction,'' 2016.

\bibitem{wu2018study}
L.~Wu, F.~Tian, T.~Qin, J.~Lai, and T.-Y. Liu, ``A study of reinforcement
  learning for neural machine translation,'' \emph{arXiv preprint
  arXiv:1808.08866}, 2018.

\bibitem{prabhavalkar2018minimum}
R.~Prabhavalkar, T.~N. Sainath, Y.~Wu, P.~Nguyen, Z.~Chen, C.-C. Chiu, and
  A.~Kannan, ``Minimum word error rate training for attention-based
  sequence-to-sequence models,'' in \emph{ICASSP}, 2018.

\bibitem{chen2022self}
C.~Chen, Y.~Hu, N.~Hou, X.~Qi, H.~Zou, and E.~S. Chng, ``Self-critical sequence
  training for automatic speech recognition,'' in \emph{ICASSP}, 2022.

\bibitem{chen2022leveraging}
C.~Chen, Y.~Hu, Q.~Zhang, H.~Zou, B.~Zhu, and E.~S. Chng, ``Leveraging
  modality-specific representations for audio-visual speech recognition via
  reinforcement learning,'' \emph{arXiv preprint arXiv:2212.05301}, 2022.

\bibitem{rennie2017self}
S.~J. Rennie, E.~Marcheret, Y.~Mroueh, J.~Ross, and V.~Goel, ``Self-critical
  sequence training for image captioning,'' in \emph{CVPR}, 2017.

\bibitem{tjandra2018sequence}
A.~Tjandra, S.~Sakti, and S.~Nakamura, ``Sequence-to-sequence asr optimization
  via reinforcement learning,'' in \emph{ICASSP}, 2018.

\bibitem{ho2020denoising}
J.~Ho, A.~Jain, and P.~Abbeel, ``Denoising diffusion probabilistic models,''
  \emph{Advances in Neural Information Processing Systems}, vol.~33, pp.
  6840--6851, 2020.

\bibitem{nichol2021improved}
A.~Q. Nichol and P.~Dhariwal, ``Improved denoising diffusion probabilistic
  models,'' in \emph{ICML}, 2021.

\bibitem{kong2020diffwave}
Z.~Kong, W.~Ping, J.~Huang, K.~Zhao, and B.~Catanzaro, ``Diffwave: A versatile
  diffusion model for audio synthesis,'' \emph{arXiv preprint
  arXiv:2009.09761}, 2020.

\bibitem{luo2021diffusion}
S.~Luo and W.~Hu, ``Diffusion probabilistic models for 3d point cloud
  generation,'' in \emph{CVPR}, 2021.

\bibitem{lu2021study}
Y.-J. Lu, Y.~Tsao, and S.~Watanabe, ``A study on speech enhancement based on
  diffusion probabilistic model,'' in \emph{2021 Asia-Pacific Signal and
  Information Processing Association Annual Summit and Conference (APSIPA
  ASC)}, 2021.

\bibitem{chen2019towards}
C.~Chen, H.-Y. Li, X.~Zhang, X.~Liu, and U.-X. Tan, ``Towards robotic picking
  of targets with background distractors using deep reinforcement learning,''
  in \emph{2019 WRC Symposium on Advanced Robotics and Automation (WRC SARA)},
  2019.

\bibitem{grondman2012survey}
I.~Grondman, L.~Busoniu, G.~A. Lopes, and R.~Babuska, ``A survey of
  actor-critic reinforcement learning: Standard and natural policy gradients,''
  \emph{IEEE Transactions on Systems, Man, and Cybernetics, Part C
  (Applications and Reviews)}, vol.~42, no.~6, 2012.

\bibitem{lu2022conditional}
Y.-J. Lu, Z.-Q. Wang, S.~Watanabe, A.~Richard, C.~Yu, and Y.~Tsao,
  ``Conditional diffusion probabilistic model for speech enhancement,'' in
  \emph{ICASSP}, 2022.

\bibitem{valentini2016investigating}
C.~Valentini-Botinhao, X.~Wang, S.~Takaki, and J.~Yamagishi, ``Investigating
  rnn-based speech enhancement methods for noise-robust text-to-speech.'' in
  \emph{SSW}, 2016.

\bibitem{garofolo1988getting}
J.~S. Garofolo, L.~F. Lamel, W.~M. Fisher, J.~G. Fiscus, and D.~S. Pallett,
  ``Getting started with the darpa timit cd-rom: An acoustic phonetic
  continuous speech database,'' \emph{National Institute of Standards and
  Technology (NIST), Gaithersburgh, MD}, vol. 107, 1988.

\bibitem{phan2020improving}
H.~Phan, I.~V. McLoughlin, L.~Pham, O.~Y. Ch{\'e}n, P.~Koch, M.~De~Vos, and
  A.~Mertins, ``Improving gans for speech enhancement,'' \emph{IEEE Signal
  Processing Letters}, 2020.

\bibitem{strauss2021flow}
M.~Strauss and B.~Edler, ``A flow-based neural network for time domain speech
  enhancement,'' in \emph{ICASSP}, 2021.

\bibitem{hsieh2020wavecrn}
T.-A. Hsieh, H.-M. Wang, X.~Lu, and Y.~Tsao, ``Wavecrn: An efficient
  convolutional recurrent neural network for end-to-end speech enhancement,''
  \emph{IEEE Signal Processing Letters}, vol.~27, pp. 2149--2153, 2020.

\bibitem{luo2019conv}
Y.~Luo and N.~Mesgarani, ``Conv-tasnet: Surpassing ideal time--frequency
  magnitude masking for speech separation,'' \emph{IEEE/ACM transactions on
  audio, speech, and language processing}, vol.~27, no.~8, pp. 1256--1266,
  2019.

\end{thebibliography}

\end{document}